# Design and characteristic study of electron blocking layer free AlInN nanowire deep ultraviolet light-emitting diodes

**Ravi Teja Velpula[1], Barsha Jain[1], Thang Ha Quoc Bui[1], Tan Thi Pham[2†], Van Thang Le[2], Hoang-Duy Nguyen[3‡], Trupti Ranjan Lenka[4], and Hieu Pham Trung Nguyen[1¥]**

[1]*Department of Electrical and Computer Engineering, New Jersey Institute of Technology, 323 Dr Martin Luther King Jr Boulevard, Newark, New Jersey, 07102*
[2]*Institute of Chemical Technology, Vietnam Academy of Science and Technology, 1 Mac Dinh Chi Street, District 1, Ho Chi Minh City 700000, Vietnam*
[3]*Ho Chi Minh City University of Technology, Vietnam National University Ho Chi Minh City, 268 Ly Thuong Kiet, Ward 14, District 10, Ho Chi Minh City 700000, Vietnam*
[4]*Department of Electronics and Communication Engineering, National Institute of Technology Silchar, Assam 788010, India*

E-mails: [†]*ptthi@hcmut.edu.vn*; [‡]*nhduy@iams.vast.vn*; [¥]*hieu.p.nguyen@njit.edu*

**Abstract:** We report on the illustration of the first electron blocking layer (EBL) free AlInN nanowire light-emitting diodes (LEDs) operating in the deep ultraviolet (DUV) wavelength region (sub-250 nm). We have systematically analyzed the results using APSYS software and compared with simulated AlGaN nanowire DUV LEDs. From the simulation results, significant efficiency droop was observed in AlGaN based devices, attributed to the significant electron leakage. However, compared to AlGaN nanowire DUV LEDs at similar emission wavelength, the proposed single quantum well (SQW) AlInN based light-emitters offer higher internal quantum efficiency without droop up to current density of 1500 A/cm$^2$ and high output optical power. Moreover, we find that transverse magnetic polarized emission is ~ 5 orders stronger than transverse electric polarized emission at 238 nm wavelength. Further research shows that the performance of the AlInN DUV nanowire LEDs decreases with multiple QWs in the active region due to the presence of the non-uniform carrier distribution in the active region. This study provides important insights on the design of new type of high performance AlInN nanowire DUV LEDs, by replacing currently used AlGaN semiconductors.

## 1. Introduction

The efficient deep ultraviolet light-emitting diodes (DUV LEDs) especially operating at sub-250 nm wavelength draw high demand in the market because of its wide range of applications such as water/air/surface disinfection [1, 2], biochemical/gas sensing [3], phototherapy [4]. Numerous efforts have been made to use AlGaN material for high efficiency DUV LEDs due to its wide bandgap optical tuning from ~ 200-365 nm. However, the performance of AlGaN DUV LEDs was fundamentally limited by the large dislocation density, the non-radiative recombination, the extremely inefficient *p*-type doping [5], low internal quantum efficiency (IQE), and low light extraction efficiency (LEE) [6-8]. Because of these limitations, AlGaN DUV LEDs have been suffering from low external quantum efficiency (EQE) and very less output optical power. Researchers also observe that the EQE is dramatically decreasing when devices are operating at sub-250nm wavelength[9-11]. For instance, the EQE of the AlGaN LEDs operating at 326 nm, 287 nm, 242 nm is recorded as 2.65%[12], 2.8%[13], and 0.012%[11] respectively. Currently, the recorded output optical power is only around 100 nW[14] for 240 nm emission wavelength, which is extremely insufficient for practical applications. The local piezoelectric field exist in the multiple quantum well (MQW) of AlGaN LEDs is responsible for the band tilting, resulting to spatial separation of electrons and holes (quantum confined stark effect, QCSE) which deteriorates the carrier radiative recombination. These phenomena lower the IQE of the LEDs[15, 16]. The IQE was improved to an extent by using nanowire[17, 18] and nanopillar[19] AlGaN LED structures, but QCSE was not suppressed completely. In addition, piezoelectric polarization and high mobility of the electrons in the LEDs lead to electron leakage into the *p*-type region and it is suppressed with the integration of electron blocking layer (EBL)[20-23]. On the other hand, if it is not well-designed, the EBL causes poor hole injection into the active region[24-26]. Furthermore, the LEE of the DUV LED devices is limited due to the switching of UV polarization from transverse electric (TE) to transverse magnetic (TM) in the DUV regime[6, 27-30].

      In order to make progress in achieving high performance DUV emitters, it is critical to identify and develop the potential of alternative UV light-emitting materials. Although $Al_xIn_{(1-x)}N$ holds high potential optical tunability in UV and visible regions, it has not been widely studied to develop UV LEDs. Recent studies have demonstrated that AlInN material provides a significant optical gain for DUV LEDs[31]. The replacement of the traditional AlGaN/GaN MQW with AlInN/GaN MQW can eliminate the defects, QCSE and strain in the UV LED[32]. Moreover, AlInN compounds can be grown on both GaN, AlN templates but growing of AlGaN compounds is inimical to be grown on GaN template[33]. Integration of AlInN as a cladding layer in InGaN/GaN MQW LED provided good carrier confinement without severe strain problem[34]. AlInN can be the alternative material for power electronic applications due its high-power figure of merit (FOM) which is more than GaN and can compete with AlGaN[35] and high-performance high electron mobility transistors (HEMTs) were demonstrated[36, 37]. In this context, we propose a novel EBL-free AlInN nanowire structure for the DUV LEDs operating at sub-250 nm wavelengths. The device structure, which contains a 200 nm thick *n*-GaN nanowire template, a 100 nm thick $n$-$Al_xIn_{(1-x)}N$, $Al_xIn_{(1-x)}N$/ $Al_yIn_{(1-y)}N$/ $Al_xIn_{(1-x)}N$ (3 nm quantum barrier (QB)/ 3 nm QW/ 3 nm QB), a 100 nm thick *p*-$Al_xIn_{(1-x)}N$ and a 10 nm thick *p*-GaN. The values of x and y are presented in the Table 1.

## 2. Results

In this work, each nanowire is considered as one LED. The schematic of the proposed structure is shown in Fig. 1(a). The simulated emission spectra of AlInN nanowire UV LEDs at room temperature is shown in Fig. 1(b). It is clearly understood that these devices can emit light in the UV A, B, and C regions. Figure 1(c) presents the emitted peak wavelength *vs* Al content (y) in the QW, which is persistent with the theoretical calculations. We have further investigated the performance of both AlInN and AlGaN based nanowire DUV LEDs and compared the simulated results. As a part of this investigation, it is necessary to know the role of the integration of EBL in DUV LEDs. We have analyzed two types of DUV LEDs that include EBL and EBL-free AlGaN nanowire LED structures. Those LEDs emit light at 320 nm and 238 nm wavelengths and the obtained results are carefully analyzed. In these simulations, 10 nm thick *p*-doped $Al_{0.57}In_{0.43}N$ and $Al_{0.98}In_{0.2}N$ layers are considered as EBLs in the 320 nm and 238 nm LEDs. Figure 2(a) shows the IQE of the LEDs that emit light at 320nm wavelength. It is seen that the integration of the EBL improves the IQE which enhances the performance of the LED. In this context, the optimized EBL can prevent the electron overflow from the active region of the LED, thus electrons overflow become almost negligible at this emission wavelength and is illustrated in Fig. 2(b). In the case of LED with EBL, consumption of the holes in the *p*-doping region is reduced due to blocking of the electrons from the active region that enhances the hole injection and is presented in Fig. 2(c). Figure 2(d) shows the IQE of the 238 nm wavelength LEDs and the integration of the EBL worsens the IQE which deteriorates the performance of the LED. In the DUV wavelengths, utilization of the EBL is responsible for poor hole injection into the active region and is shown in Fig. 2(f). As a result, the electron overflow increases from the active region and it can be understood from Fig. 2(e), which is similar to recent reported studies[25]. Hence, EBL-free LEDs are desirable in the DUV region. In this regard, we demonstrated that AlInN nanowire structures offer a perfect approach to overcome such aforementioned problem.

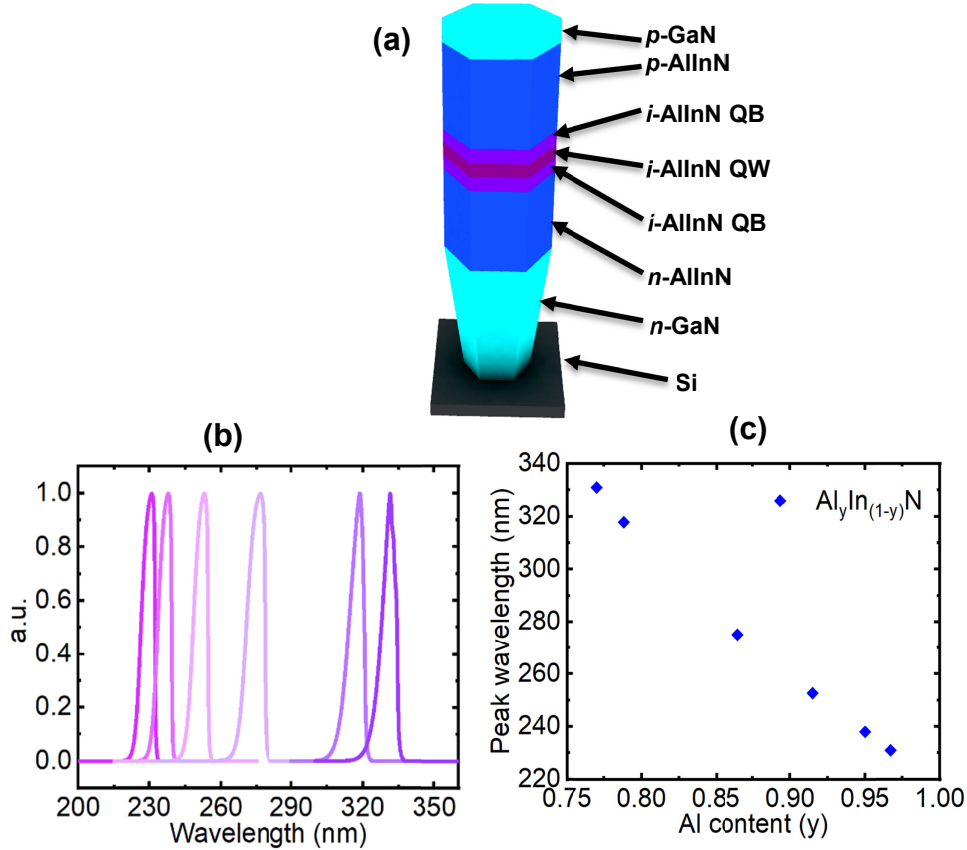

**Fig. 1**. (a) Schematic diagram of the AlInN UV LED. (b) Multiple peak emissions varied from 230 nm to 331 nm wavelengths. (c) Emission peak wavelength *vs* Al content in AlInN active region.

The EBL-free AlInN and AlGaN nanowire DUV LEDs are simulated at 238 nm emission wavelength. For this analysis, AlGaN LEDs with SQW, 3QWs and 5QWs and AlInN with SQW in their active regions are considered. The results of the 238 nm emission wavelength LEDs are shown in Fig. 3. Figure 3(a) shows the IQE of those LEDs. The IQE of the AlGaN LEDs is increasing with number of QWs in the active region but the devices are suffering from IQE droop thus decreases the device performance. The IQE droop is increasing with the injected current in the AlGaN nanowire DUV LEDs and this trend is clearly understood from Fig. 3(a). Besides, AlInN nanowire LEDs with SQW shows no efficiency droop attributed to the enhanced device performance. The electron leakage from the active region plays dominant role in the IQE droop and the electron leakage current density is decreasing with number of QWs in the active region of AlGaN nanowire LEDs. However, significant amount of the electron leakage can be

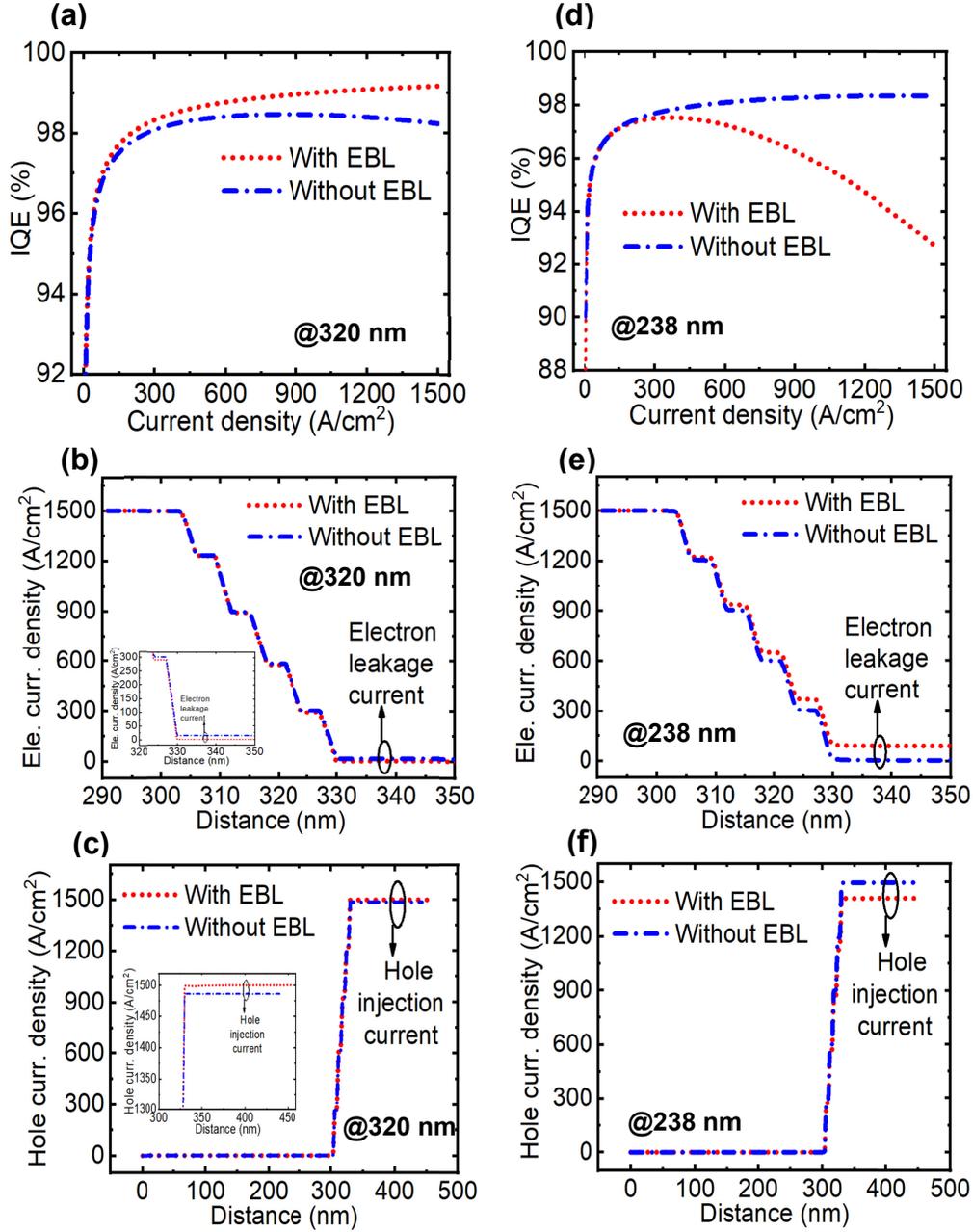

**Fig. 2**. Simulated (a, d) Internal quantum efficiency (IQE), (b, e) Electron current density, (c, f) Hole current density of with/without EBL AlGaN nanowire UV LEDs at 320 nm and 238 nm emission wavelengths.

observed even in the AlGaN nanowire LEDs with 3 and 5 QWs. In contrary, the AlInN nanowire LEDs with SQW have almost negligible electron leakage from the active region. These phenomena can be understood from Fig. 3(b). The droop free IQE of AlInN LEDs is responsible for the high radiative recombination thus it leads to obtain more output optical power as compared to other devices. This phenomenon can be clearly seen from Fig. 3(c). Figure 3(d) shows the I-V characteristics of AlInN LEDs and the I-V characteristics of AlGaN

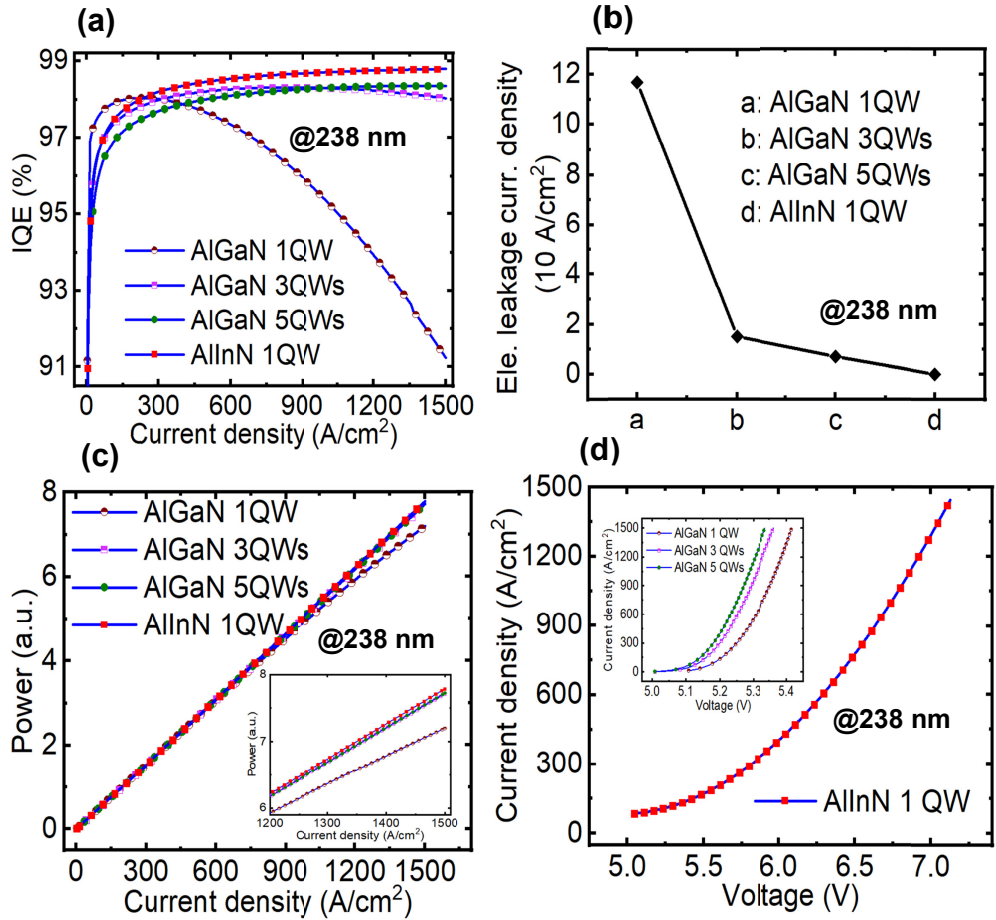

**Fig. 3**. Simulated (a) Internal quantum efficiency. (b) Electron leakage current density. (c) L-I characteristics. (d) I-V characteristics of AlGaN and AlInN LEDs at 238 nm emission wavelength.

LEDs are shown in the inset figure. AlInN LEDs have higher turn-on voltage as compared to others due to presence of high resistance in the device which is generated due to presence of more Al content in the device than other devices at respective emission wavelengths and is presented in Table 1.

We have further investigated the performance of the AlInN nanowire DUV LEDs by simulating with SQW, 3 QWs and 5 QWs in the active region at 238 nm emission wavelength. Figure 4(a) shows the IQE of those AlInN nanowire LEDs. We have observed that the SQW LED shows better IQE than MQW LED devices due to inhomogeneity of carrier distribution in the MQW active region[38]. Moreover, it has higher output power as compared to MQW LEDs and is illustrated in Fig. 4(b). Figures 4(c) and 4(d) show the carrier concentration of the SQW and 5 QWs AlInN nanowire LEDs. The non-uniform carrier distribution can be understood from Fig. 4(d). Radiative recombination phenomena of these LEDs are presented in Figs. 4(e) and 4(f). In MQW LED case, the radiative recombination rates become very non-uniform and dominant radiative recombination can be observed in the QW which is near to the *p*-region[39, 40]. This arises due to poor carrier transport from one

QW to another. The calculated total radiative recombination rate of SQW LED is $9.25 \times 10^{28}/cm^3 s$, which is higher than 5 QWs LED.

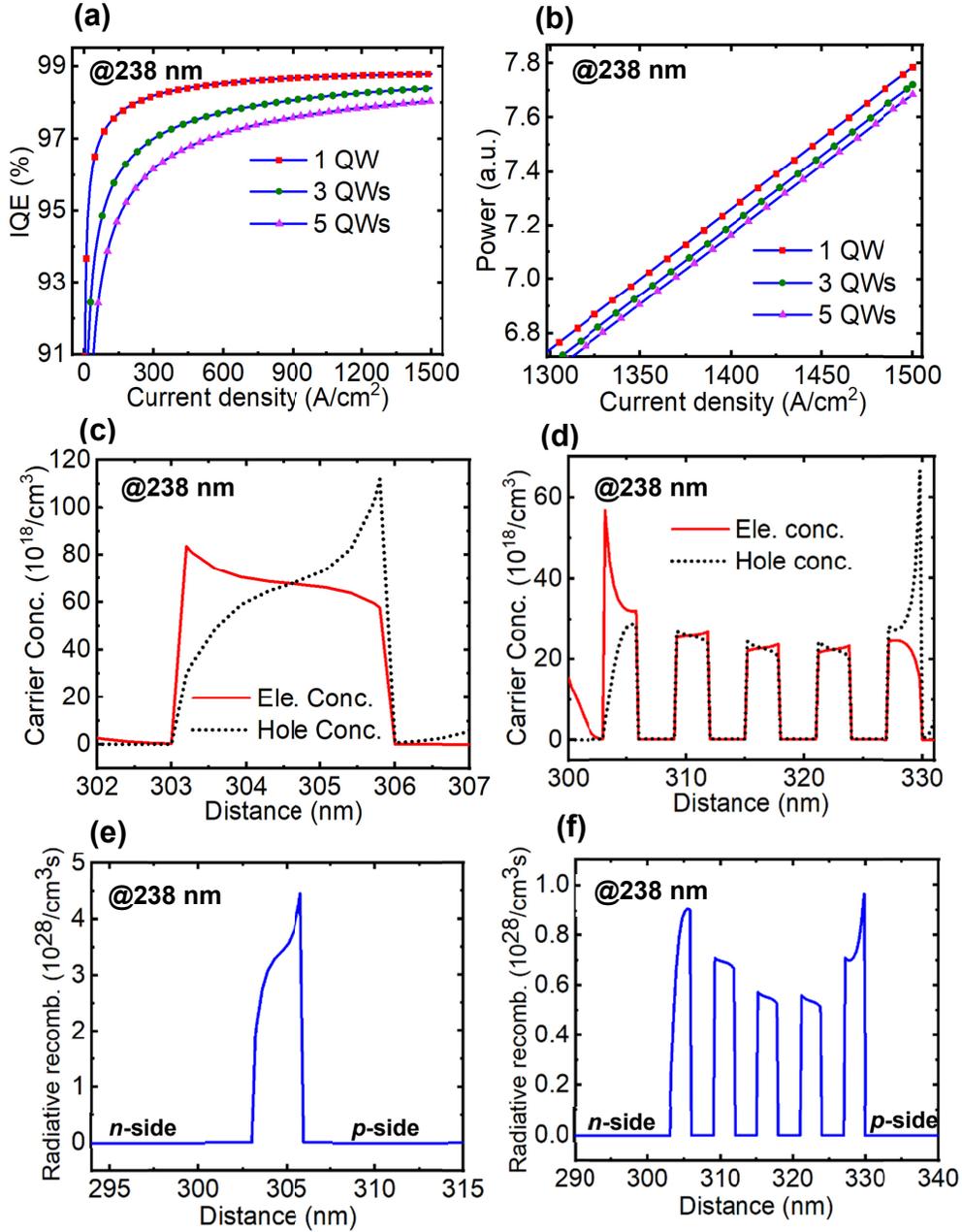

**Fig. 4**. Simulated (a) Internal quantum efficiency. (b) L-I characteristics for AlInN LEDs with SQW, 3 QWs and 5 QWs. Carrier concentration of AlInN (c) SQW LED. (d) 5 QWs LED. Radiative recombination of AlInN (e) SQW LED. (f) 5 QWs LED. at 238 nm emission wavelength.

The total radiative recombination of 5 QWs LED is $9.18 \times 10^{28}/cm^3 s$. MQW AlInN LEDs have less non-uniformity in their carrier distribution as compared to InGaN/AlGaN devices[39, 40] due to strong carrier confinement with negligible electron leakage from the active region.

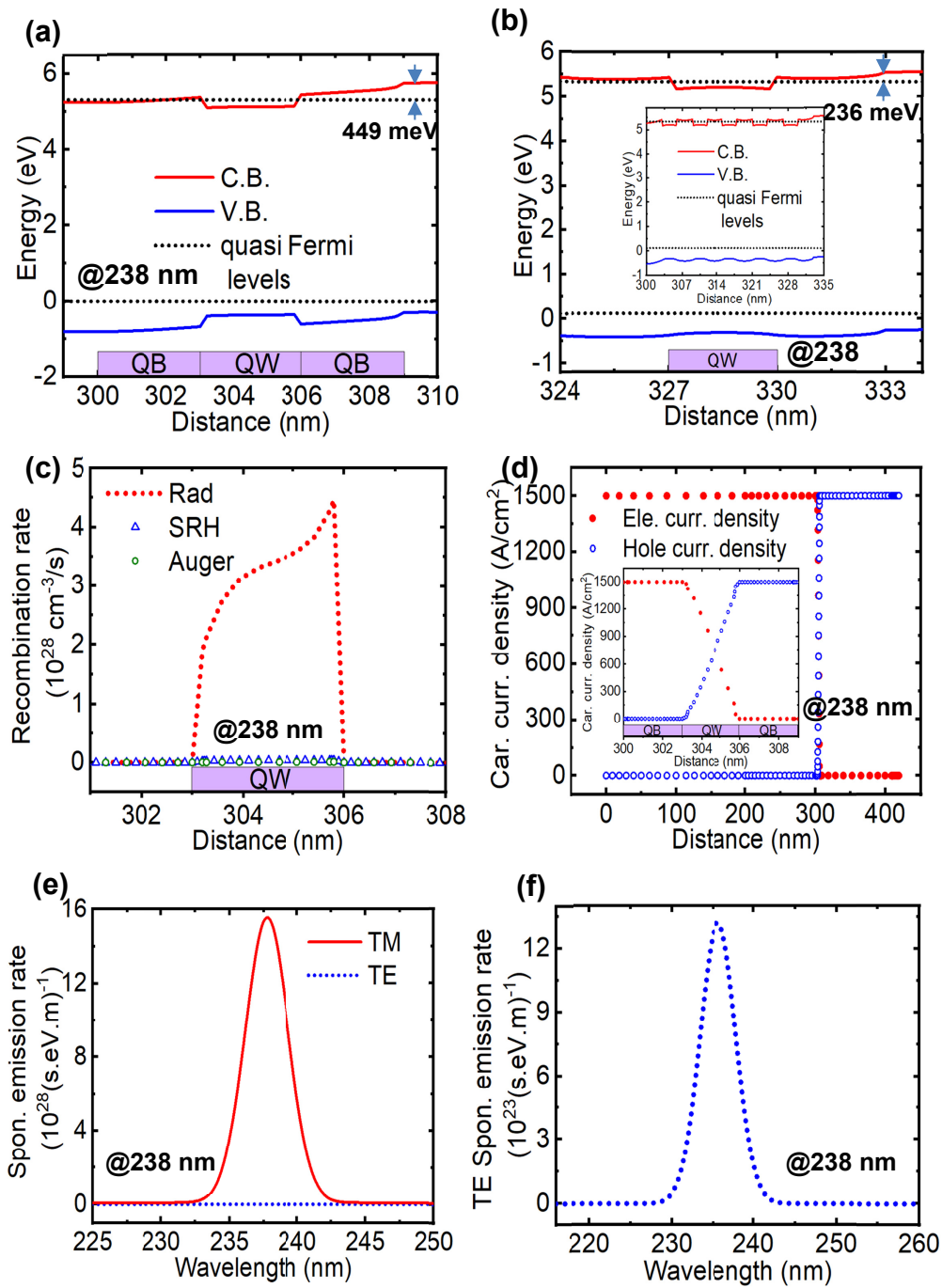

Fig. 5. Simulated (a) E-B diagram of active region of AlInN SQW LED. (b) E-B diagram of the last QW in active region of 5 QWs AlGaN LED (5QWS are shown in inset figure.) (c) Recombination rate (d) Carrier current density. (e) Spontaneous emission rate (TE/TM). (f) TE spontaneous emission rate of AlInN SQW LED at 238 nm emission wavelength.

In order to better understand the mechanism of improvements in SQW AlInN nanowire LED at 238 nm emission wavelength, the energy band (E-B) diagram is investigated. Figure 5(a) shows the E-B diagram of active region of SQW AlInN nanowire LED. The last QW in active region of 5QWs AlGaN LED is shown in Fig. 5(b) and complete active region is presented in inset figure. It is clearly understood that there is an energy band bending at the interface of the last QB and the *p*-type region due to the polarization effect, which may increase the electron leakage from the active region into the *p*-type region. The effective potential height for electron in the conduction band for AlInN LED and AlGaN LED are 449 meV, 236 meV respectively and are shown in Figs. 5(a) and 5(b). As a result, AlGaN LED has less electrons in the QW and significant electron leakage into *p*-type region, which consumes holes in *p*-type region. Because of the lower carriers in the active region AlGaN LED has lower radiative recombination. On the other hand, AlInN LED has high effective potential height at the interface of the QB and the *p*-type region, which prevents the electron leakage and provides better carrier recombination in the active region. The carrier recombination of the AlInN LEDs is shown in Fig. 5(c). It is observed that the radiative recombination is the dominant mechanism which is ~2 orders more than that of SRH and Auger recombination. The best E-B structure and dominant radiative recombination of AlInN LEDs made the devices have almost negligible carrier current density outside of the active region, which can be observed from Fig. 5 (d). Thus, the proposed structure is a self EBL structure with no electron leakage into the *p*-type region. Here, we can able to almost block all the electrons in the active region without the help of the EBL, which enhances the hole injection efficiency in the active region[24, 25]. In addition, AlInN LEDs have almost negligible hole current density as well before the QW, which supports that most of the injected carriers are radiatively recombining. As a result, AlInN LEDs have high IQE without any droop.

The polarization properties of the emission from AlInN UV nanowire LEDs at room temperature were also analyzed. The TM and TE polarized emissions are defined as the electric fields parallel (**E**//**c**) and perpendicular (**E** ⊥ **c**) to c-axis, respectively. The calculation was performed at injection current of 1500 A/cm$^2$. Illustrated in Figs. 5(e) and 5(f), the UV light emission is predominantly TM polarized which is about ~ 5 orders stronger than that of TE polarized emission. Similar trend of polarization for LEDs using AlGaN in the same UV wavelength regime are also reported in other reports[41-44]. This result plays important role in the design of surface emitting UV LEDs using AlInN compounds to achieve high LEE. The LEE of the UV LEDs can be improved with surface plasmon[45] and modified surface structures[46].

## 3. Discussions

III-nitride LEDs suffer from inefficient carrier confinement in the active regions, leading to significant electron leakage. In this regard, the integration of an optimized EBL in the LED structure can mitigate this problem. However, the integration of the EBL in the LED not only blocks the electron leakage but also lowers the hole injection into the active region causing poor hole transport problem. In addition, the integration of EBL increases lattice mismatch, local electrical field and may allow electrons to escape from QW and spill over EBL[24]. Hence, the LED without EBL, with efficient carrier confinement capability, negligible carrier leakage from active region is the optimal choice. From the above results, it is evident that the AlInN nanowire LEDs are the optimal choice and potential alternative of AlGaN LEDs in the DUV light-emitter development. The main challenge with AlInN DUV LEDs is the high resistance which causes LEDs to have high turn on voltage. Further optimization in terms of device structure with lower resistance of the device and composition will be performed in order to achieve high power DUV light emission using AlInN nanowire structures.

TABLE 1: Parameters of the 238 nm wavelength AlInN and AlGaN nanowire LEDs.

| Layer | Thickness (nm) | Al content in AlInN LED | Al content in AlGaN LED |
|---|---|---|---|
| $n$ - GaN | 200 | - | - |
| $n$ - $Al_xIn_{(1-x)}N$ | 100 | 0.975 | 0.91 |
| $i$ - $Al_xIn_{(1-x)}N$ | 3 | 0.975 | 0.91 |
| $i$ - $Al_yIn_{(1-y)}N$ | 3 | 0.95 | 0.84 |
| $i$ - $Al_xIn_{(1-x)}N$ | 3 | 0.975 | 0.91 |
| $p$ - $Al_xIn_{(1-x)}N$ | 100 | 0.975 | 0.91 |
| $p$ - GaN | 10 | - | - |

## 4. Summary

We have successfully demonstrated the EBL free AlInN nanowire LEDs operating in the DUV region with negligible efficiency droop at room temperature. We have also observed that SQW AlInN DUV LEDs have better performance in terms of IQE and output optical power compared to other AlGaN based DUV LEDs. Additionally, the advantage of our proposed SQW AlInN device is also from the epitaxial growth of such SQW structure which is relatively simple to fabricate as compared to MQW LED structures in term of dislocation density and homogeneity of In distribution in the AlInN epilayers. The electron and hole overflows were not observed with these SQW AlInN DUV emitters. The devices exhibit stable emission with strong TM polarized emission. The device performance can be further improved by engineering the device structure.

**Acknowledgements**
This research is supported by New Jersey Institute of Technology, the National Science Foundation grant EEC-1560131, and funded by Vietnam National Foundation for Science and Technology Development (NAFOSTED) under grant number 103.03-2017.312.